\documentclass{article} % For LaTeX2e
\usepackage{arxiv}
\usepackage{natbib}   % arxiv.sty 不带，ICLR 的 sty 原来带
\usepackage{times}

\usepackage{amsmath,amsfonts,bm}

\def\eqref#1{equation~\ref{#1}}
\def\1{\bm{1}}

\DeclareMathAlphabet{\mathsfit}{\encodingdefault}{\sfdefault}{m}{sl}
\SetMathAlphabet{\mathsfit}{bold}{\encodingdefault}{\sfdefault}{bx}{n}

\usepackage{hyperref}
\usepackage{url}

\usepackage{amsmath}
\usepackage{algorithm}
\usepackage{algorithmic}
\usepackage{amssymb}

\usepackage{booktabs}   % \toprule \midrule \bottomrule
\usepackage{graphicx}   % \resizebox
\usepackage[table]{xcolor}
\usepackage{multirow}
\newcommand{\ci}[1]{\ensuremath{^{\scalebox{0.9}{$\scriptstyle\pm #1$}}}}
\title{MoGeFlow: Flowing Through Motion Codebook Geometry for Text-to-Motion Generation}
\author{
Pengcheng Fang\textsuperscript{1, 3},
Tengjiao Sun\textsuperscript{1, 3},
Xiaoyu Zhan\textsuperscript{2,3},
Hansung Kim\textsuperscript{1},
Xiaohao Cai\textsuperscript{1},
Dongjie Fu\textsuperscript{3, \textdagger},
\\[0.8em]
\textsuperscript{1}University of Southampton \quad
\textsuperscript{2}Nanjing University \quad 
\textsuperscript{3}MOGO AI \quad
\\[0.4em]
\textsuperscript{\textdagger}Corresponding author.
}

\begin{document}

\maketitle

\begin{abstract}
Vector-quantized tokenizers have become the dominant interface for text-to-motion generation, and the priors built on them treat motion codes as unordered categorical labels.
We show that this discards structure the tokenizer already learned.
A motion code embedding is a decoder-bound movement prototype, and learned motion codebooks carry measurable, decoder-causal geometry: code distances track the distances between the movements those codes decode to, the alignment collapses under shuffled controls, and displacement in code space predicts how far the frozen decoder moves.
Geometry of this kind calls for a different generative object---a continuous trajectory through the code space rather than a sequence of decisions over a vocabulary.
We introduce \textbf{MoGeFlow}, a text-conditioned flow over part-structured motion-code frames.
Its tokenizer quantizes each body-part group separately, so every decoder input is a point on a combinatorial lattice of decoder-bound states, and the flow regresses onto that lattice.
Because the decoder responds smoothly to code-space displacement, generated states are decoded as they are, and the detail the flow places between codes reaches the decoder instead of being rounded away.
Quantization therefore serves generation entirely at training time, shaping the space rather than gating the samples: under matched conditions the lattice-shaped target beats both the unquantized latents it is built from and a matched variational latent space.
MoGeFlow sets a new state of the art in text--motion alignment on HumanML3D, leads Top-3 retrieval and MultiModal Distance on KIT-ML, and on the large-scale MotionMillion benchmark surpasses a 7B-parameter model in retrieval with an order of magnitude fewer parameters.
Quantization shapes the space; the flow generates through it. \href{https://github.com/PengchengFang-cs/MoGeFlow}{github.com/PengchengFang-cs/MoGeFlow}
\end{abstract}

\section{Introduction}

Text-to-motion generation synthesizes human motion from natural-language descriptions, a core capability for animation, games, virtual reality, and embodied agents whose action models increasingly build on generative motion priors~\citep{fang2026pfd}.
Continuous methods run diffusion or flow models directly on motion trajectories or autoencoder latents~\citep{tevet2023mdm, zhang2024motiondiffuse, chen2023mld}, while token-based methods compress motion with a vector-quantized tokenizer~\citep{oord2017neural} and learn a text-conditioned prior over code sequences~\citep{zhang2023t2mgpt, jiang2023motiongpt, guo2024momask, fu2026mogo}.
The token interface is compact, admits strong sequence models, and ships with a frozen decoder---but its prior is trained as if the codebook were an unordered vocabulary: predict the next index, or recover the masked ones.
A categorical loss over indices charges the same price for a code that decodes to a nearly identical movement and for one that decodes to something unrelated.

That price is wrong only if the codebook has geometry to lose, and whether it does is not a matter of taste.
A code embedding is a prototype that a frozen decoder consumes to produce movement, so \emph{do nearby embeddings decode to nearby movements} is a measurable property of a trained tokenizer, and the answer dictates what a prior should generate: over an unordered vocabulary, categorical prediction is the natural interface; over a geometry, generation can move continuously.

We measure it, and the answer is unambiguous.
On a frozen part-structured tokenizer, code-embedding distance tracks the distance between the movements those codes decode to in every body-part group; shuffling the code-to-movement correspondence destroys the alignment, ruling out usage frequency and index identity; and replacing a code with progressively farther neighbors moves the frozen decoder's output progressively farther, while an equal-length displacement in a random direction moves it $1.8$ to $7.2$ times less.
A learned motion codebook is therefore decoder-causal geometry rather than a vocabulary with arbitrary labels, and the property is learned: an untrained tokenizer of the same architecture scores far lower.

\textbf{MoGeFlow} is the generative consequence of that measurement.
Its tokenizer quantizes each of six data-derived body-part groups separately, so the reachable decoder inputs form a combinatorial product lattice, and a text-conditioned flow transports noise onto that lattice, supervising displacement in the very space where distance predicts decoded change.
At inference the terminal state is handed to the frozen decoder unquantized: its offset from the lattice is not isotropic but concentrated along local codebook directions, the directions along which the decoder responds most, so decoding it as it stands is better than rounding it first.
We test that inversion of the usual reading of quantization rather than assume it---holding the flow and the tokenizer fixed, the lattice target outperforms the same tokenizer's pre-quantization latents and a matched variational latent space, and re-quantizing terminal states is worse than decoding them directly.

We contribute \emph{motion codebook geometry} as a measurable, decoder-causal property of learned tokenizers; a flow over a part-structured code lattice that generates through it and decodes without quantization; and the finding that quantization contributes training-time shaping of the target space rather than an interface sampling must return to.
We evaluate on HumanML3D~\citep{guo2022generating}, KIT-ML~\citep{plappert2016kit}, and MotionMillion.

\section{Related Work}

\noindent\textbf{Text-to-motion generation.}
Continuous methods generate motion with diffusion or flow objectives, in trajectory space~\citep{tevet2023mdm, zhang2024motiondiffuse} or in a variational latent space~\citep{petrovich2022temos, chen2023mld}, and recent flow-based systems extend this template~\citep{guan2026flowcomotion, li2026motionhiflow}.
Token-based methods compress motion with vector-quantized tokenizers~\citep{oord2017neural} and learn categorical priors over code indices, whether autoregressive~\citep{zhang2023t2mgpt, jiang2023motiongpt, pinyoanuntapong2024bamm}, masked~\citep{guo2024momask, pinyoanuntapong2024mmm, guo2025snapmogen}, or hierarchical over residual levels~\citep{fu2026mogo, fu2025hirqct}.
Two of them set the operating points we compare against: SALAD~\citep{hong2025salad} denoises in the continuous latent space of a skeleto-temporal VAE and holds the strongest published retrieval alignment, while MoGenTS~\citep{yuan2024mogents} quantizes joints on a spatio-temporal grid and predicts indices over it.
The two families share one blind spot from opposite sides: categorical priors discard the geometry of the space they predict over, while latent-space generators work in continuous spaces with no decoder-bound anchors.
MoGeFlow keeps the structural bias of the former and the continuity of the latter.

\noindent\textbf{Continuous generation for quantized representations.}
Flow matching transports a simple prior to a data distribution with a learned vector field~\citep{lipman2023flow}; for discrete targets the open question is which space the flow should live in.
Language and vision either diffuse over embeddings and round back to the vocabulary, or drop quantization and generate continuous tokens directly~\citep{li2022diffusionlm, dieleman2022cdcd, chen2023analogbits, tschannen2024givt, li2024mar}.
Closest in mechanism, the VQ-regularized variant of latent diffusion~\citep{rombach2022ldm} diffuses in the continuous latent \emph{before} the quantization layer, so the quantizer acts as a latent regularizer rather than a sampling interface.
MoGeFlow moves in the opposite direction: it regresses onto the quantized lattice itself, keeping the decoder-bound anchors that pre-quantization diffusion gives up---and where this line adopts a continuous space as a convenience and never measures it, we treat a motion-code state as a decoder-bound movement prototype and test that geometry against decoded motion.

\noindent\textbf{Structured and controllable motion generation.}
Part-aware generation coordinates motion across body regions~\citep{zou2024parco}, and a parallel line broadens conditioning and control interfaces~\citep{cohan2024condmdi, dai2024motionlcm, zhang2025motionduet, zhan2026umo, fu2026mogo, fang2026anchorroute}.
MoGeFlow adds no conditioning modality; its part structure is the statistical PartVQ partition of KV-Control~\citep{kvcontrol}, whose groups are discovered from motion statistics and corrected for kinematic-chain integrity, and what it contributes over that interface is representational---a product lattice, a measurement of its geometry, and a prior that generates through it.

%%%%%%%%%%%%%%%%%%%%%%%%%%%%%%%%%%%
\section{Method}

\begin{figure}[t]
    \centering
    \includegraphics[width=1.0\linewidth]{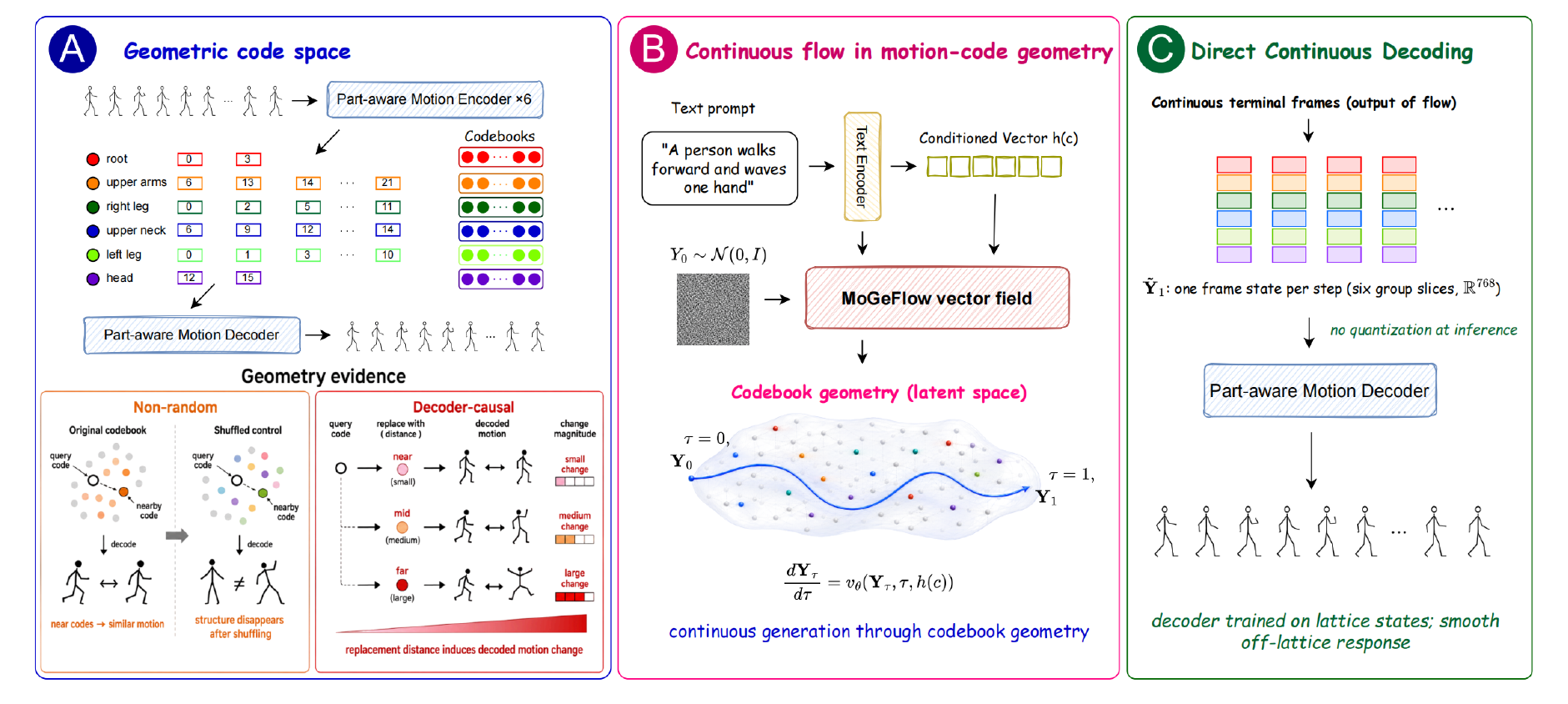}
    \caption{
Overview of \textbf{MoGeFlow}.
A frozen part-structured tokenizer encodes motion into one code embedding per data-derived joint group---root, upper arms, right leg, upper neck, left leg, and head---so every decoder-bound frame concatenates one codebook vector per group and lies on a combinatorial product lattice.
These groups are statistically discovered rather than manually predefined left/right or upper/lower body partitions.
MoGeFlow learns a text-conditioned continuous flow over these structured frame states and hands terminal states directly to the frozen motion decoder; no quantization is applied at inference.
}
    \label{fig:mogeflow_overview}
\end{figure}

A part-structured vector-quantized tokenizer admits only finitely many decoder input frames: one codebook vector per joint group, concatenated.
Finite, however, is not small.
The admissible frames form a Cartesian product over groups, and in our tokenizer this product contains roughly $4.4\times10^{12}$ distinct decoder-bound states per frame.
MoGeFlow treats this lattice, rather than the code indices that address it, as the generative target (Figure~\ref{fig:mogeflow_overview}).

Given a text description $c$, MoGeFlow transports noise to structured frame states and decodes the result:
\[
    \mathbf{Y}_0 \sim \mathcal{N}(0,I),
    \qquad
    \tilde{\mathbf{Y}}_1 = \Phi_\theta^1(\mathbf{Y}_0;c),
    \qquad
    \hat{x}=D_\phi\bigl(\mathcal{W}^{-1}(\tilde{\mathbf{Y}}_1)\bigr),
\]
where $\Phi_\theta^1$ is the terminal map of a text-conditioned flow over motion-code frames, $\mathcal{W}$ is the fixed per-dimension standardization of code space defined in Section~\ref{sec:method_tokenizer}, and $D_\phi$ is the frozen motion decoder.
No quantization separates the flow from the decoder: terminal states are decoded directly, a choice the measured geometry of the code space licenses and Section~\ref{sec:experiments} validates against re-quantized decoding.
The rest of this section proceeds from interface to prior: we formalize the tokenizer and the lattice it induces, measure the geometry this lattice carries, and construct the text-conditioned flow that generates through it.

\subsection{A Part-Structured Tokenizer and its Code Lattice}
\label{sec:method_tokenizer}

MoGeFlow generates through a part-structured tokenizer that is trained once and then frozen; $\phi$ denotes the union of its parameters---part encoders, codebooks, and decoder---and no subsequent stage updates $\phi$.
Let $x\in\mathcal{D}$ be a motion sequence of $F$ frames.
The skeleton is divided into $P$ data-derived joint groups $\mathcal{J}=\{\mathcal{J}_1,\ldots,\mathcal{J}_P\}$, which may share boundary joints.
The grouping follows KV-Control's PartVQ analysis: per-joint activations and lag-aware joint correlations define a statistical joint similarity, hierarchical clustering proposes candidate groups, and a kinematic-chain correction keeps major limb chains intact.
With $P=6$ the groups are descriptively named root, upper arms, right leg, upper neck, left leg, and head after their dominant joints; the names identify learned groups rather than a hand-designed split, and Appendix~\ref{app:joint_groups} lists their exact composition.

For each group $p$, a temporal convolutional encoder compresses the group's channels to a quarter-rate latent sequence $r_{t,p}(x)=E_\phi^p(x_{\mathcal{J}_p})_t\in\mathbb{R}^{d}$ with $t=1,\ldots,T$ and $T=F/4$, and each group owns a codebook $\mathcal{E}_p=\{e_{p,1},\ldots,e_{p,K}\}$ with $e_{p,k}\in\mathbb{R}^{d}$.
Each group is quantized exactly once: the tokenizer has no residual levels, so a code frame is a single concatenation of $P$ codebook vectors.
Quantization selects the nearest code independently for every group and step,
\[
    z_{t,p}(x)
    =
    \arg\min_{k\in\{1,\ldots,K\}}
    \|r_{t,p}(x)-e_{p,k}\|_2^2,
    \qquad
    q_{t,p}(x)=e_{p,z_{t,p}(x)},
\]
and we write $Z(x)=\{z_{t,p}(x)\}_{t,p}$ for the id grid and $\operatorname{Emb}_{\mathcal{E}}(Z)=\{e_{p,z_{t,p}}\}_{t,p}$ for the embeddings of any id grid.

MoGeFlow's unit of generation is the whole-body code frame that concatenates the $P$ quantized group embeddings at each step, $\mathbf{y}_{1,t}(x)=[q_{t,1}(x);\cdots;q_{t,P}(x)]\in\mathbb{R}^{Pd}$.
Collecting frames defines the tokenizer map
\[
    F_\phi(x)
    =
    \bigl(\mathbf{y}_{1,1}(x),\ldots,\mathbf{y}_{1,T}(x)\bigr)
    =:
    \mathbf{Y}_1(x)
    \in
    \mathbb{R}^{T\times Pd}.
\]
The frozen decoder consumes exactly this representation: $\hat{x}=D_\phi(\mathbf{Y}_1)$, and it accepts any point of $\mathbb{R}^{T\times Pd}$, quantized or not.
Every frame of every training target lies on the product lattice $\mathcal{S}=\mathcal{E}_1\times\cdots\times\mathcal{E}_P\subset\mathbb{R}^{Pd}$ of size $|\mathcal{S}|=K^P$, embedded by concatenation; with $K{=}128$ and $P{=}6$ this yields roughly $4.4\times10^{12}$ decoder-bound states per frame.
Group-specific codebooks preserve the specialization discovered by the partition and frame-level concatenation composes them into coordinated whole-body states, so the lattice factorizes over groups yet grows exponentially with $P$.
Because every input the decoder sees during tokenizer training is a lattice point, reconstruction training concentrates decoder competence on and around $\mathcal{S}$---the property the next subsection measures and the flow later exploits.

\noindent\textbf{Working coordinates.}
Group codebooks differ in scale, so the flow is defined in standardized code space.
Let $\mu_p,\sigma_p\in\mathbb{R}^{d}$ be the per-dimension mean and standard deviation of the entries of $\mathcal{E}_p$, both fixed once the tokenizer is frozen, and let
\begin{align*}
    \mathcal{W}(\mathbf{y})
    &=
    \bigl[(\mathbf{y}_{1}-\mu_1)\oslash\sigma_1;\;\cdots;\;(\mathbf{y}_{P}-\mu_P)\oslash\sigma_P\bigr],
    \\
    \mathcal{W}^{-1}(\mathbf{u})
    &=
    \bigl[\sigma_1\odot \mathbf{u}_{1}+\mu_1;\;\cdots;\;\sigma_P\odot \mathbf{u}_{P}+\mu_P\bigr],
\end{align*}
acting frame-wise, where $\mathbf{y}_p$ denotes the block of $\mathbf{y}$ belonging to group $p$.
$\mathcal{W}$ is an invertible group-wise affine map, so it carries the lattice to a standardized lattice $\mathcal{W}(\mathcal{S})$ without changing which states are reachable.
Sections~\ref{sec:method_flow} and~\ref{sec:method_sampling} work in these coordinates and the decoder is always applied through $\mathcal{W}^{-1}$; the diagnostics below report both metrics.

Our instantiation uses $P{=}6$, $K{=}128$, and $d{=}128$, so each code frame summarizes four motion frames as a $768$-dimensional lattice state; Appendix~\ref{app:implementation} gives the tokenizer's training recipe.

\subsection{Motion Codebook Geometry Diagnostics}
\label{sec:method_diagnostics}

The lattice of Section~\ref{sec:method_tokenizer} is a useful domain for continuous generation only if it carries a property its construction does not guarantee: distances in code space must predict differences in decoded motion.
We measure this on the frozen tokenizer before building the prior, with three diagnostics computed on a held-out corpus $\mathcal{D}'$; none of them trains or tunes any parameter.

For group $p$ and code $k$, collect the occurrences $\Omega_{p,k}=\{(x,t): x\in\mathcal{D}',\, z_{t,p}(x)=k\}$ and keep the codes with sufficient support, $\mathcal{K}_p^+=\{k: |\Omega_{p,k}|\geq n_{\min}\}$ with $n_{\min}=20$; active-code counts are reported alongside the results.
Averaging the group's normalized motion features over a code's occurrences yields that code's local motion prototype,
\[
    \bar{g}_{p,k}
    =
    \frac{1}{|\Omega_{p,k}|}
    \sum_{(x,t)\in\Omega_{p,k}}
    g_p(x,t),
    \qquad k\in\mathcal{K}_p^+,
\]
where $g_p(x,t)$ denotes the normalized motion features of group $p$ over the temporal support of tokenizer step $t$.
A second, decoder-side prototype substitutes the code into real contexts and averages what the frozen decoder then produces at that step (Appendix~\ref{app:geometry_protocol}).
Each active code therefore carries an embedding $e_{p,k}$ that the decoder consumes, together with prototypes for the motion it stands for and the motion it produces.

\noindent\textbf{Metric alignment.}
The two representations induce two metrics on the codebook, $D^{\mathcal{E}}_{p}(k,k')=\|e_{p,k}-e_{p,k'}\|_2$ and $D^{\mathcal{M}}_{p}(k,k')=\|\bar{g}_{p,k}-\bar{g}_{p,k'}\|_2$, and we quantify their rank agreement per group,
\[
    \rho_p
    =
    \operatorname{Spearman}
    \left(
        \{D^{\mathcal{E}}_{p}(k,k')\}_{k<k'},
        \{D^{\mathcal{M}}_{p}(k,k')\}_{k<k'}
    \right),
    \qquad k,k'\in\mathcal{K}_p^+ .
\]

\noindent\textbf{Shuffled control.}
High alignment could in principle reflect usage statistics rather than learned structure.
We therefore recompute $\rho_p$ after randomly permuting the association between code embeddings and motion prototypes while preserving usage statistics, which removes any correspondence the tokenizer learned while leaving the marginal usage of each code untouched.

\noindent\textbf{Replacement causality.}
Alignment is correlational; generation depends on how the decoder responds.
Starting from an encoded sequence $Z(x)$, we replace a single code $z_{t,p}$ with a code $k'$ at a controlled embedding distance, decode both sequences, and measure the induced motion change,
\[
    \Delta(k')
    =
    d_{\mathrm{dec}}
    \left(
        D_\phi(\operatorname{Emb}_{\mathcal{E}}(Z)),
        D_\phi(\operatorname{Emb}_{\mathcal{E}}(Z^{t,p\rightarrow k'}))
    \right),
\]
where $d_{\mathrm{dec}}$ is the mean absolute per-channel deviation from the unmodified decode, reported for the whole body, for the replaced group, and locally at the replaced step.
If the geometry matters to the decoder, $\Delta$ must grow with replacement distance.

These diagnostics probe the lattice at its own points---codes and single-code replacements---and the same protocol extends off it: Section~\ref{sec:method_sampling} measures where the terminal states of the flow sit relative to their nearest codes and which directions their offsets take, and Section~\ref{sec:experiments} compares decoding those states directly against rounding them first.

\subsection{Text-Conditioned Flow Matching in Codebook Space}
\label{sec:method_flow}

The tokenizer induces a target distribution over standardized code frames: for $(x,c)\sim p_{\mathrm{data}}$, the map $\mathcal{W}\circ F_\phi$ pushes data onto the standardized lattice, and we write $\mu_1(\cdot\mid c)$ for the resulting conditional distribution, supported on $\mathcal{W}(\mathcal{S})^T$.
MoGeFlow learns to transport a Gaussian base $\mu_0=\mathcal{N}(0,I)$ to $\mu_1(\cdot\mid c)$ with a text-conditioned ODE over structured motion-code frames,
\[
    \frac{d\mathbf{Y}_\tau}{d\tau}
    =
    v_\theta(\mathbf{Y}_\tau,\tau,h(c)),
    \qquad
    \tau\in[0,1],
\]
where $v_\theta$ is a neural vector field and $h(c)$ is the text representation: the per-token feature sequence of a frozen CLIP ViT-B/32 text encoder, projected linearly to the model width, with no part of the encoder finetuned.
Text enters through joint attention over these token features, while the pooled sentence embedding modulates every block together with the flow time $\tau$ through adaptive layer normalization.
Rotary positions index motion tokens by time step while all text tokens share a constant position, so text participates order-free at the attention level and word order is carried inside the frozen encoder's features.
For classifier-free guidance, captions are dropped with probability $0.1$ during training, and $h(\varnothing)$ is the encoding of the empty caption by the same frozen encoder.

We train the vector field with conditional flow matching~\citep{lipman2023flow}.
With a clean endpoint $\mathbf{Y}_1=\mathcal{W}(F_\phi(x))$, noise $\mathbf{Y}_0\sim\mathcal{N}(0,I)$ and $\tau\sim\mathcal{U}(0,1)$, the rectified path and its target velocity are
\[
    \mathbf{Y}_\tau=(1-\tau)\mathbf{Y}_0+\tau\mathbf{Y}_1,
    \qquad
    u_\tau=\mathbf{Y}_1-\mathbf{Y}_0,
\]
and the training objective is the masked regression
\[
    \mathcal{L}_{\mathrm{flow}}
    =
    \mathbb{E}_{(x,c),\mathbf{Y}_0,\tau}
    \left[
    \frac{1}{\sum_t m_t}
    \sum_{t=1}^{T}
    m_t
    \left\|
        v_\theta(\mathbf{Y}_\tau,\tau,h(c))_t
        -
        u_{\tau,t}
    \right\|_2^2
    \right],
\]
with $m_t\in\{0,1\}$ the valid-frame mask.
The supervision is thus vector-valued in the space where local code distances are meaningful, rather than categorical over labels: an error is charged in proportion to how far it moves the state in code space, which by Section~\ref{sec:method_diagnostics} tracks how far it moves the decoder's output.

The vector field is a text-conditioned Transformer over structured frame states.
Each temporal position is one frame token, assembled from per-part projections, and the flow time enters only through modulation:
\[
    \mathbf{h}^{(0)}_t=\bigl[\,W^{\mathrm{in}}_1\,\mathrm{LN}_1(\mathbf{y}_{\tau,t,1});\;\cdots;\;W^{\mathrm{in}}_P\,\mathrm{LN}_P(\mathbf{y}_{\tau,t,P})\,\bigr]\in\mathbb{R}^{D},
\]
where $D=P\cdot d_{\mathrm{part}}$.
The flow time and the pooled sentence embedding $\bar{h}(c)$ together form the modulation vector
\[
    \psi=\mathrm{MLP}\bigl(\mathrm{sinemb}(\tau)\bigr)+W^{\mathrm{txt}}\bar{h}(c).
\]
Following the multimodal diffusion transformer design of~\citet{esser2024sd3}, the backbone applies $L_d$ double-stream blocks, in which motion and text streams attend jointly under separate modulation, followed by $L_s$ single-stream blocks over the concatenated sequence; attention uses rotary positions (time-step index for motion tokens, a constant index for text tokens), with no additive positional or timestep embedding.
Writing $\mathbf{h}^{(L)}_t$ for the final hidden state of frame $t$, each part owns a modulated output head that predicts the clean endpoint,
\[
    \hat{\mathbf{Y}}_{1,\theta}(\mathbf{Y}_\tau,\tau,h(c))_{t,p}
    =
    W^{\mathrm{out}}_p\Bigl[\bigl(1+\gamma_p(\psi)\bigr)\odot\mathrm{LN}(\mathbf{h}^{(L)}_t)+\beta_p(\psi)\Bigr],
\]
and the vector field is formed from it in closed form, $v_\theta(\mathbf{Y}_\tau,\tau,h(c))=\bigl(\hat{\mathbf{Y}}_{1,\theta}(\mathbf{Y}_\tau,\tau,h(c))-\mathbf{Y}_\tau\bigr)/\max(1-\tau,\epsilon)$ with $\epsilon=10^{-4}$; the loss is taken in velocity space, so the endpoint error it charges carries a $(1-\tau)^{-2}$ weight.
Temporal attention models dynamics across code frames while each frame token retains the part-specific structure needed for whole-body coordination.

\subsection{Objective, Sampling, and Direct Decoding}
\label{sec:method_sampling}

MoGeFlow trains with the flow-matching loss alone, $\mathcal{L}_{\mathrm{MoGeFlow}}=\mathcal{L}_{\mathrm{flow}}$, with no auxiliary code-classification or terminal losses: quantization influences training only through where the targets $\mathbf{Y}_1$ lie.

At inference we sample $\mathbf{Y}^{0}\sim\mathcal{N}(0,I)$ and integrate the learned ODE on a uniform grid $\tau_n=n/N$, writing $\mathbf{Y}^{n}$ for the state at $\tau_n$ and $\Delta\tau=1/N$:
\begin{align*}
    \mathbf{Y}^{n+1}
    &=
    \mathbf{Y}^{n}
    +
    \Delta\tau\,
    \tilde{v}(\mathbf{Y}^{n},\tau_n),
    \\
    \tilde{v}(\mathbf{Y},\tau)
    &=
    v_\theta(\mathbf{Y},\tau,h(\varnothing))
    +
    w
    \left[
        v_\theta(\mathbf{Y},\tau,h(c))
        -
        v_\theta(\mathbf{Y},\tau,h(\varnothing))
    \right],
\end{align*}
where $w$ is the classifier-free guidance scale.

The terminal state is returned to code space and handed to the frozen decoder unchanged,
\[
    \hat{x}
    =
    D_\phi\bigl(\mathcal{W}^{-1}(\mathbf{Y}^{N})\bigr).
\]
No quantization intervenes.
Terminal states concentrate near the lattice without landing on it: on generated motions the median per-group offset to the nearest code is $0.68$ of half the median nearest-neighbor spacing, and $26\%$ of frame--group states lie beyond it (Appendix~\ref{app:geometry_protocol}).
These offsets are not isotropic: they carry three and a half times more energy along the local codebook directions than an equally long displacement in a random direction, and Section~\ref{sec:method_diagnostics} measured those directions to be the ones the decoder responds to most.
Re-quantizing the terminal state discards them.

%%%%%%%%%%%%%%%%%%%%%%%%%%%%%%%%%%%%%%%%%%%%%%%%%%%%%%%%%%%%%%%%%

\section{Experiments}
\label{sec:experiments}

We ask three questions: how MoGeFlow compares with recent text-to-motion methods, whether the frozen codebooks carry the geometry the method assumes, and what the design owes to the codebook interface as opposed to capacity.

\subsection{Experimental Setup}
\label{sec:exp_setup}

\noindent\textbf{Datasets.}
We evaluate on HumanML3D~\citep{guo2022generating}, KIT-ML~\citep{plappert2016kit}, and the large-scale MotionMillion~\citep{fan2025motionmillion}, each under its standard protocol and evaluator.

\noindent\textbf{Metrics.}
We report R-Precision (Top-1/2/3), FID, MultiModal Distance, Diversity, and MultiModality, with ``$\pm$'' the half-width of the 95\% confidence interval over 20 replications on HumanML3D and KIT-ML.
All MoGeFlow results condition on the ground-truth motion length; no length predictor is used, and Appendix~\ref{app:reported_exp_details} gives the remaining protocol details.

\noindent\textbf{Models.}
Main benchmark results use MoGeFlow-L; ablations report the S/B/L variants.

\subsection{Main Text-to-Motion Results}
\label{sec:main_results}

\begin{table*}[t]
\centering
\scriptsize
\setlength{\tabcolsep}{2pt}
\renewcommand{\arraystretch}{1.08}
\caption{
Quantitative comparisons with recent representative and state-of-the-art methods on HumanML3D (upper half) and KIT-ML (lower half).
``$\pm$'' is the half-width of the 95\% confidence interval under the repeat-20 protocol; \textbf{bold} and \underline{underline} mark the best and second-best generated-motion results in R-Precision, FID and MultiModal Distance; real motion is shown as a reference and is not ranked.
}
\label{tab:main_results}
\setlength{\tabcolsep}{1.5pt}
\begin{tabular}{llccccccc}
\toprule
\multirow{2}{*}{Methods} & \multirow{2}{*}{Venue}
& \multicolumn{3}{c}{R-Precision $\uparrow$}
& \multirow{2}{*}{FID $\downarrow$}
& \multirow{2}{*}{MultiModal Dist $\downarrow$}
& \multirow{2}{*}{Diversity $\rightarrow$}
& \multirow{2}{*}{MultiModality $\uparrow$} \\
\cmidrule(lr){3-5}
& & Top1 & Top2 & Top3 & & & & \\
\midrule

\rowcolor{gray!20}
\multicolumn{9}{l}{\textit{On the HumanML3D dataset.}} \\
Real motion & -
& $0.511\ci{.003}$ & $0.703\ci{.003}$ & $0.797\ci{.002}$
& $0.002\ci{.000}$ & $2.974\ci{.008}$ & $9.503\ci{.065}$ & $-$ \\
\midrule
T2M-GPT & CVPR'23
& $0.492\ci{.003}$ & $0.679\ci{.002}$ & $0.775\ci{.002}$
& $0.141\ci{.005}$ & $3.121\ci{.009}$ & $9.761\ci{.081}$ & $1.831\ci{.048}$ \\
ReMoDiffuse & ICCV'23
& $0.510\ci{.005}$ & $0.698\ci{.006}$ & $0.795\ci{.004}$
& $0.103\ci{.004}$ & $2.974\ci{.016}$ & $9.018\ci{.075}$ & $1.795\ci{.043}$ \\
MoMask & CVPR'24
& $0.521\ci{.002}$ & $0.713\ci{.002}$ & $0.807\ci{.002}$
& $0.045\ci{.002}$ & $2.958\ci{.008}$ & $-$ & $1.241\ci{.040}$ \\
BAMM & ECCV'24
& $0.525\ci{.002}$ & $0.720\ci{.003}$ & $0.814\ci{.003}$
& $0.055\ci{.002}$ & $2.919\ci{.008}$ & $9.717\ci{.089}$ & $1.687\ci{.051}$ \\
MoGenTS & NeurIPS'24
& $0.529\ci{.003}$ & $0.719\ci{.002}$ & $0.812\ci{.002}$
& $\underline{0.033}\ci{.001}$ & $2.867\ci{.006}$ & $9.570\ci{.077}$ & $-$ \\
Light-T2M & AAAI'25
& $0.511\ci{.003}$ & $0.699\ci{.002}$ & $0.795\ci{.002}$
& $0.040\ci{.002}$ & $3.002\ci{.008}$ & $-$ & $1.670\ci{.061}$ \\
IRG-MotionLLM & ECCV'26
& $0.535\ci{.002}$ & $0.725\ci{.002}$ & $0.820\ci{.002}$
& $0.242\ci{.006}$ & $2.785\ci{.006}$ & $9.900\ci{.094}$ & $-$ \\
EnergyMoGen & CVPR'25
& $0.526\ci{.003}$ & $0.718\ci{.003}$ & $0.815\ci{.002}$
& $0.176\ci{.006}$ & $2.931\ci{.007}$ & $9.500\ci{.091}$ & $2.205\ci{.041}$ \\
SALAD & CVPR'25
& $\underline{0.581}\ci{.003}$ & $\underline{0.769}\ci{.003}$ & $\underline{0.857}\ci{.002}$
& $0.076\ci{.002}$ & $\underline{2.649}\ci{.009}$ & $9.696\ci{.096}$ & $1.751\ci{.062}$ \\
MoMask++ & NeurIPS'25
& $0.528\ci{.003}$ & $0.718\ci{.003}$ & $0.811\ci{.002}$
& $0.072\ci{.003}$ & $2.912\ci{.008}$ & $-$ & $1.227\ci{.046}$ \\
MotionHiFlow & CVPR'26 
& $0.563\ci{.003}$ & $0.754\ci{.003}$ & $0.843\ci{.003}$
& $\textbf{0.032}\ci{.002}$ & $2.691\ci{.009}$ & $9.504\ci{.071}$ & $-$ \\
\midrule
MoGeFlow & -
& $\textbf{0.589}\ci{.002}$ & $\textbf{0.784}\ci{.002}$ & $\textbf{0.874}\ci{.002}$
& $0.048\ci{.003}$ & $\textbf{2.600}\ci{.008}$ & $9.914\ci{.081}$ & $1.216$ \\

\midrule
\rowcolor{gray!20}
\multicolumn{9}{l}{\textit{On the KIT-ML dataset.}} \\
Real motion & -
& $0.424\ci{.005}$ & $0.649\ci{.006}$ & $0.779\ci{.006}$
& $0.031\ci{.004}$ & $2.788\ci{.012}$ & $11.080\ci{.097}$ & $-$ \\
\midrule
T2M-GPT & CVPR'23
& $0.416\ci{.006}$ & $0.627\ci{.006}$ & $0.745\ci{.006}$
& $0.514\ci{.029}$ & $3.007\ci{.023}$ & $10.860\ci{.094}$ & $1.570\ci{.039}$ \\
ReMoDiffuse & ICCV'23
& $0.427\ci{.014}$ & $0.641\ci{.004}$ & $0.765\ci{.055}$
& $0.155\ci{.006}$ & $2.814\ci{.012}$ & $10.800\ci{.105}$ & $1.239\ci{.028}$ \\
MoMask & CVPR'24
& $0.433\ci{.007}$ & $0.656\ci{.005}$ & $0.781\ci{.005}$
& $0.204\ci{.011}$ & $2.779\ci{.022}$ & $-$ & $1.131\ci{.043}$ \\
BAMM & ECCV'24
& $0.438\ci{.009}$ & $0.661\ci{.009}$ & $0.788\ci{.005}$
& $0.183\ci{.013}$ & $2.723\ci{.026}$ & $11.008\ci{.094}$ & $1.609\ci{.065}$ \\
MoGenTS & NeurIPS'24
& $0.445\ci{.006}$ & $0.671\ci{.006}$ & $0.797\ci{.005}$
& $\underline{0.143}\ci{.004}$ & $2.711\ci{.024}$ & $10.918\ci{.090}$ & $-$ \\
Light-T2M & AAAI'25
& $0.444\ci{.006}$ & $0.670\ci{.007}$ & $0.794\ci{.005}$
& $0.161\ci{.009}$ & $2.746\ci{.016}$ & $-$ & $1.005\ci{.036}$ \\
IRG-MotionLLM & ECCV'26
& $0.445\ci{.005}$ & $0.681\ci{.003}$ & $0.781\ci{.004}$
& $0.432\ci{.013}$ & $2.740\ci{.017}$ & $11.115\ci{.086}$ & $-$ \\
EnergyMoGen & CVPR'25
& $0.436\ci{.006}$ & $0.651\ci{.006}$ & $0.772\ci{.006}$
& $0.495\ci{.020}$ & $2.861\ci{.020}$ & $11.060\ci{.101}$ & $1.256\ci{.024}$ \\
SALAD & CVPR'25
& $0.477\ci{.006}$ & $\underline{0.711}\ci{.005}$ & $\underline{0.828}\ci{.005}$
& $0.296\ci{.012}$ & $2.585\ci{.016}$ & $11.097\ci{.095}$ & $1.004\ci{.040}$ \\
MotionHiFlow & CVPR'26 
& $\textbf{0.482}\ci{.005}$ & $0.704\ci{.005}$ & $0.825\ci{.005}$
& $\textbf{0.135}\ci{.007}$ & $\underline{2.552}\ci{.014}$ & $10.894\ci{.117}$ & $-$ \\
\midrule
MoGeFlow & -
& $\underline{0.479}\ci{.007}$ & $\textbf{0.712}\ci{.006}$ & $\textbf{0.842}\ci{.004}$
& $0.145\ci{.005}$ & $\textbf{2.499}\ci{.016}$ & $10.724\ci{.093}$ & $0.947$ \\
\bottomrule
\end{tabular}
\end{table*}

\begin{figure*}[t]
    \centering
    \includegraphics[width=0.85\linewidth]{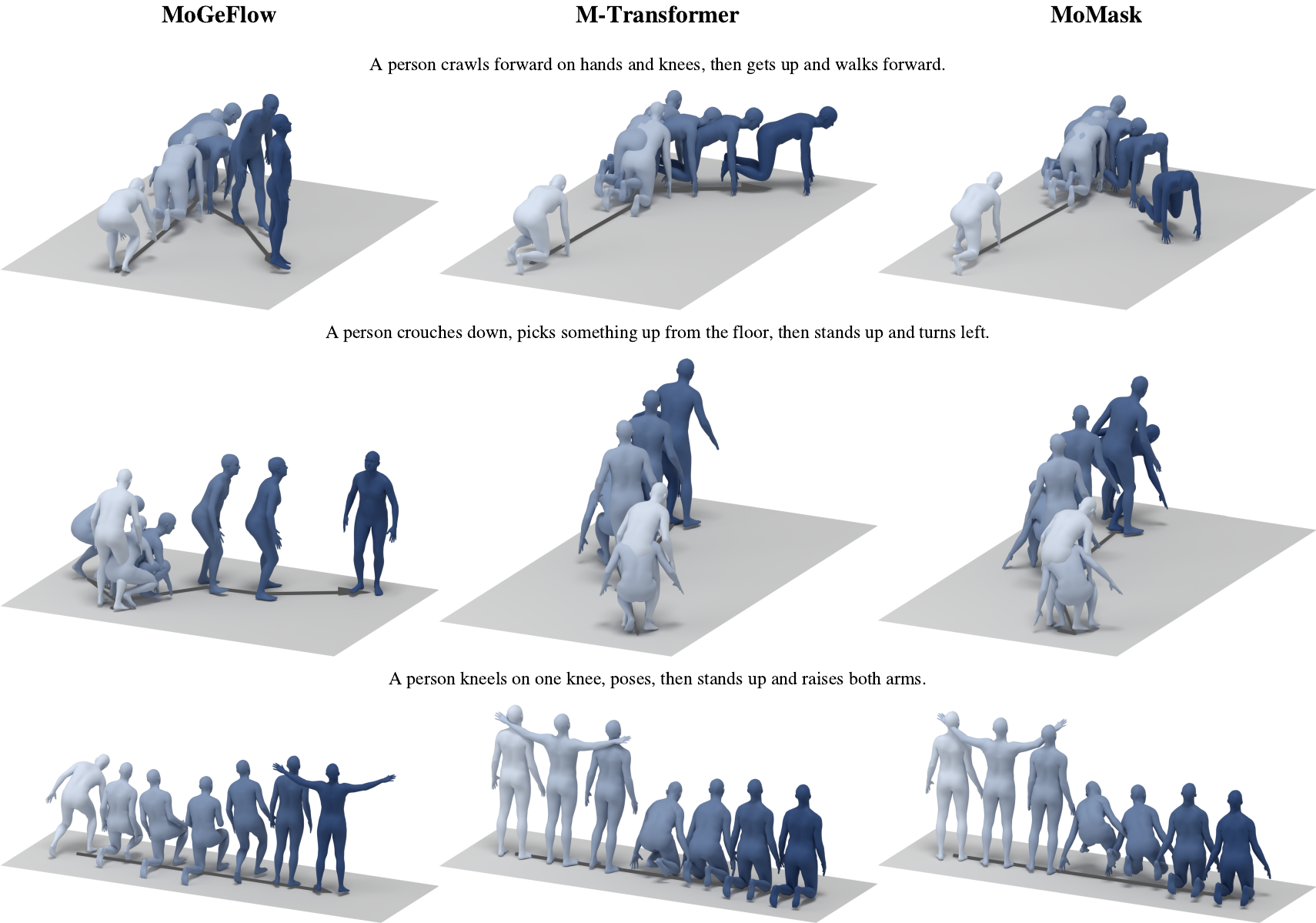}
    \caption{
    Qualitative comparison on multi-stage text prompts.
    Each row shows one text condition, and each column shows one generated motion from MoGeFlow, M-Transformer~\citep{guo2022generating}, and MoMask~\citep{guo2024momask}.
    Motions are visualized by overlaying seven SMPL body meshes on the ground plane, shaded from light to dark as time runs from the first keyframe to the last; arrows indicate the global movement direction when visible.
    MoGeFlow better preserves the ordered action phases specified by the text.
    }
    \label{fig:qualitative_comparison}
\end{figure*}

Table~\ref{tab:main_results} compares MoGeFlow against diffusion and flow models in trajectory or latent space~\citep{zhang2023remodiffuse, zeng2025lightt2m, zhang2025energymogen, li2026motionhiflow}, token-based categorical priors~\citep{zhang2023t2mgpt, guo2024momask, pinyoanuntapong2024bamm, yuan2024mogents} including the MoMask++ prior of~\citet{guo2025snapmogen}, the language-model prior of~\citet{li2025irgmotionllm}, and SALAD~\citep{hong2025salad}, whose real-motion reference rows we adopt.
MoGeFlow leads text--motion alignment on both benchmarks---best R-Precision at every rank on HumanML3D, best Top-3 on KIT-ML, and best MultiModal Distance on both---at an FID of $0.048$ on HumanML3D and $0.145$ on KIT-ML.
Figure~\ref{fig:qualitative_comparison} compares generated motions on multi-stage prompts.

\begin{table*}[t]
\raggedright
\caption{
Left: MotionMillion benchmark comparison under its evaluation protocol, against ScaMo~\citep{lu2025scamo} and MotionMillion-7B~\citep{fan2025motionmillion}.
Right: decoder-causal geometry diagnostic---a code is displaced towards a near, mid-range, or far codebook neighbor (\textbf{bold}; mean code distance $13.0$, $64.6$, $195.9$) or by the same distance in a random direction, and each cell reports how far the frozen decoder's output moves, written as neighbor\,/\,random.
}
\label{tab:motionmillion_and_replacement}
\footnotesize
\setlength{\tabcolsep}{2pt}
\renewcommand{\arraystretch}{1.05}

\begin{minipage}[t]{0.46\textwidth}
\centering
\textbf{(a) MotionMillion}

\vspace{1mm}
\begin{tabular}{lcccc}
\toprule
Method & FID$\downarrow$ & R@1$\uparrow$ & R@2$\uparrow$ & R@3$\uparrow$ \\
\midrule
ScaMo & 89.0 & 0.67 & 0.81 & 0.87 \\
MotionMillion-7B & \textbf{10.3} & 0.79 & 0.90 & 0.94 \\
MoGeFlow-L & \underline{12.1} & \textbf{0.93} & \textbf{0.99} & \textbf{0.99} \\
\bottomrule
\end{tabular}
\end{minipage}
\hspace{0.02\textwidth}
\begin{minipage}[t]{0.50\textwidth}
\centering
\textbf{(b) Code replacement}

\vspace{1mm}
\begin{tabular}{lccc}
\toprule
Type & Body $\Delta$ & Group $\Delta$ & Local $\Delta$ \\
\midrule
near & \textbf{0.0062} / 0.0028 & \textbf{0.0106} / 0.0034 & \textbf{0.062} / 0.015 \\
mid  & \textbf{0.0284} / 0.0121 & \textbf{0.0614} / 0.0166 & \textbf{0.567} / 0.079 \\
far  & \textbf{0.0684} / 0.0370 & \textbf{0.1532} / 0.0531 & \textbf{1.285} / 0.309 \\
\bottomrule
\end{tabular}
\end{minipage}
\end{table*}

The pattern holds at scale.
On MotionMillion (Table~\ref{tab:motionmillion_and_replacement}(a)) MoGeFlow attains the best retrieval at every rank against both ScaMo and a 7B-parameter baseline, with an order of magnitude fewer parameters.

\subsection{Do Motion Codebooks Carry Geometry?}
\label{sec:geometry_results}

\begin{table}[t]
\centering
\caption{
Left: motion codebook geometry diagnostics on held-out HumanML3D motions. Group-specific codebook distances align with local motion-prototype distances in the raw code metric, in the per-dimension standardized metric the flow is trained under, and against prototypes obtained by decoding each code with the frozen decoder, while shuffled controls collapse it to a mean of $+0.014$; bootstrap 95\% confidence intervals on the correlation columns have width $\le 0.030$ and each group codebook contains 128 entries.
Right: training-target ablation on HumanML3D, where the flow recipe and evaluation are held fixed and the regression target is changed to the unquantized encoder latents of the same tokenizer, or to the posterior mean of a matched variational tokenizer.
}
\label{tab:geometry_diagnostics}
\footnotesize
\setlength{\tabcolsep}{1pt}
\renewcommand{\arraystretch}{1.05}
\begin{minipage}[t]{0.57\textwidth}
\centering
\textbf{(a) Codebook geometry}

\vspace{1mm}
\begin{tabular}{lccccc}
\toprule
Group
& Codes
& $\rho$ $\uparrow$
& $\rho$ std. $\uparrow$
& $\rho$ decoded $\uparrow$
& $\rho$ shuffled \\
\midrule
root       & 123 & 0.962 & 0.964 & 0.962 & $-0.037$ \\
upper arms & 126 & 0.852 & 0.854 & 0.964 & $+0.059$ \\
right leg  & 106 & 0.779 & 0.836 & 0.603 & $-0.033$ \\
upper neck & 111 & 0.903 & 0.917 & 0.942 & $+0.135$ \\
left leg   & 113 & 0.629 & 0.692 & 0.595 & $+0.002$ \\
head       &  99 & 0.803 & 0.804 & 0.938 & $-0.039$ \\
\midrule
Mean       & 113.0 & 0.822 & 0.845 & 0.834 & $+0.014$ \\
\bottomrule
\end{tabular}
\end{minipage}
\hfill
\begin{minipage}[t]{0.41\textwidth}
\centering
\textbf{(b) Training-target space}

\vspace{1mm}
\begin{tabular}{llcc}
\toprule
Setting & Target & R@3$\uparrow$ & FID$\downarrow$ \\
\midrule
L width & codebook & \textbf{0.874} & \textbf{0.048} \\
 & encoder latent & 0.853 & 0.082 \\
\midrule
B width & codebook & \textbf{0.864} & \textbf{0.054} \\
 & variational latent & 0.846 & 0.088 \\
\bottomrule
\end{tabular}
\end{minipage}
\end{table}

Tables~\ref{tab:geometry_diagnostics}(a) and~\ref{tab:motionmillion_and_replacement}(b) test the premise the method rests on.
Code-embedding distance agrees with motion-prototype distance in every group in the raw code metric, and the agreement collapses under the shuffled control to a mean of $+0.014$, so it reflects learned correspondence rather than usage statistics or index identity.
The same correlation computed against decoded prototypes---each code substituted into real contexts and passed through the frozen decoder---is $0.834$: code distance predicts the distance between the movements the decoder actually produces.
It is also decoder-causal: farther codebook substitutions move the decoder's output farther, at whole-body, group, and local scales alike.
An equal-length displacement in a random direction moves it $1.8$ to $7.2$ times less: the decoder is not merely smooth, its response concentrates along the directions the codebook spans.
Three further controls sharpen the claim.
Re-initialising the encoder and codebooks at random and repeating the diagnostic with that tokenizer's own assignments drops the mean correlation to $0.24$, so the alignment is learned rather than a by-product of nearest-neighbor assignment.
Applied to a generic residual-VQ tokenizer the protocol holds at its base level but not across residual levels (Appendix~\ref{app:geometry_protocol}), so most of the space a residual-VQ flow regresses over is only weakly decoder-aligned---consistent with its weaker joint alignment--fidelity in Table~\ref{tab:key_ablations}.
The same correlations computed in the standardized metric the flow is trained under are equal or higher (Table~\ref{tab:geometry_diagnostics}(a)), so the geometry holds in the space the model optimizes.
Together this motivates the design: generation should move continuously through this geometry, and terminal states can be decoded without rounding (protocol in Appendix~\ref{app:geometry_protocol}).

\subsection{Ablation Studies}
\label{sec:key_ablations}

\begin{table}[t]
\centering
\caption{
Key ablations on HumanML3D, and the effect of decoding terminal states directly instead of rounding them to their nearest codes.
}
\label{tab:key_ablations}
\setlength{\tabcolsep}{2pt}
\begin{tabular}{lccccc}
\toprule
Variant 
& Prior 
& Code Interface 
& Params 
& R@3 $\uparrow$ 
& FID $\downarrow$ \\
\midrule
Generic RVQ + flow (stage grid)
& flow
& generic RVQ
& 677M
& $0.833\ci{.004}$
& $0.062\ci{.005}$ \\

Generic RVQ + flow (summed latent)
& flow
& generic RVQ
& 694M
& $0.854\ci{.005}$
& $0.094\ci{.005}$ \\

Masked categorical
& categorical
& part-specific
& 121M
& $0.826\ci{.003}$
& $0.083\ci{.002}$ \\

MoGeFlow-S
& flow
& part-specific
& 173M
& $0.856\ci{.003}$
& $0.077\ci{.003}$ \\

MoGeFlow-B
& flow
& part-specific
& 308M
& $\underline{0.864}\ci{.002}$
& $\underline{0.054}\ci{.002}$ \\

MoGeFlow-L
& flow
& part-specific
& 690M
& $\textbf{0.874}\ci{.002}$
& $\textbf{0.048}\ci{.003}$ \\
\midrule
\multicolumn{4}{l}{\emph{Decoding of the same terminal states}} & & \\
Nearest re-quantization & -- & -- & -- & $0.873\ci{.002}$ & $0.058\ci{.003}$ \\
Direct continuous & -- & -- & -- & $\textbf{0.874}\ci{.002}$ & $\textbf{0.048}\ci{.003}$ \\
\bottomrule
\end{tabular}
\end{table}

Table~\ref{tab:key_ablations} separates what the design owes to the interface from what it owes to capacity.

At a matched parameter budget (677M and 694M against 690M) and the same flow recipe, neither generic residual-VQ variant reaches alignment and fidelity together, while the part-structured lattice does.
Continuous flow alone is therefore not the active ingredient; applying it over this lattice is.
The objective matters on the same lattice as well: the masked categorical prior is nearest in size to MoGeFlow-S, and the flow at that scale is ahead of it by $0.030$ R@3 at a lower FID.
Width improves both axes; what the matched-budget arms isolate is the interface rather than the parameter count.

Two design choices remain.
Table~\ref{tab:key_ablations} also decodes the same terminal states both ways: direct decoding improves FID from $0.058$ to $0.048$ while retrieval is unchanged, so what the terminal offsets carry is worth keeping (Sec.~\ref{sec:method_sampling}).
Changing only the training target (Table~\ref{tab:geometry_diagnostics}(b)), the identical flow aimed at the same tokenizer's pre-quantization latents, or at a variational latent space of matched architecture, is weaker on both axes.
Quantization therefore earns its place at training time: among the target spaces we tested, the lattice-shaped one is the strongest (protocol in Appendix~\ref{app:ablation_protocol}).

\section{Conclusion}
\label{sec:conclusion}

MoGeFlow generates through motion codebook geometry: code distance predicts how far the frozen decoder moves, so a text-conditioned flow over the part-structured lattice can be decoded without quantization, and it leads text--motion alignment on HumanML3D, KIT-ML, and MotionMillion.
The lattice comes from a tokenizer we do not train; learning it jointly with the flow is the natural next step.

\subsection*{Ethics statement}

This work uses only publicly released motion-capture corpora under their
original licences and access conditions, and collects no new human-subject
data. The motions are distributed as de-identified skeletal and parametric
sequences, and we neither recover nor attempt to recover the identity of the
performers. Text-conditioned motion synthesis produces movement that reads as
plausible without being grounded in any recorded person, and the alignment and
fidelity we report are measured by learned evaluators rather than by human
judgement; the model is therefore not validated for any setting in which the
realism of a motion carries consequences, such as biomechanical, ergonomic or
clinical analysis. Generated motion can also be used to animate avatars in
ways that misrepresent real people, a risk this method shares with the
text-to-motion literature it builds on and one that any deployment would need
to address.

\subsection*{Reproducibility statement}

The frozen tokenizer and the product lattice it induces are defined in
Section~\ref{sec:method_tokenizer}, the geometry diagnostics and their
estimators in Section~\ref{sec:method_diagnostics}, and the text-conditioned
flow, its objective, and the sampler in
Section~\ref{sec:method_flow} and Section~\ref{sec:method_sampling}.
Appendix~\ref{app:geometry_protocol} gives the full diagnostic protocol,
including the prototype construction on both the encoder and the decoder side,
the shuffled, untrained-tokenizer, norm-matched and support-matched controls,
the terminal off-lattice measurements, and the resampling used for every
interval we report. Appendix~\ref{app:joint_groups} lists the exact joint
composition of the six data-derived groups together with the boundary channels
they share. Appendix~\ref{app:benchmark_eval} states the evaluation protocol
and the number of replications behind every reported interval,
Appendix~\ref{app:ablation_protocol} defines each ablation arm and what is held
fixed across it, and Appendix~\ref{app:implementation} gives the optimizer,
schedule, precision, sampler settings and guidance scale. All three corpora are
public and are used under the standard evaluation protocol of each benchmark.

\subsection*{AI use statement}

Large language models were used as a general-purpose writing and coding
assistant. Their use was limited to editing prose for grammar and concision,
formatting \LaTeX{} tables, and drafting boilerplate portions of the
experiment-running and result-aggregation scripts. The research idea, the
measurement of codebook geometry, the design of the generative model, the
experimental design, and the interpretation of all results are the authors'
own. Every number we report for our own models comes from runs executed and
verified by the authors, and the authors take full responsibility for the entire content
of the paper, including any text that an assistant helped edit.

\bibliography{iclr2026_conference}
\bibliographystyle{iclr2026_conference}

% \appendix
% \section{Appendix}
\appendix

\section{Details for Reported Experiments}
\label{app:reported_exp_details}

This appendix provides implementation details for the experiments reported in Sec.~\ref{sec:experiments}.

\subsection{Benchmark Evaluation}
\label{app:benchmark_eval}

\noindent\textbf{HumanML3D and KIT-ML.}
For HumanML3D and KIT-ML, we follow the standard text-to-motion evaluation protocol with 20 replications and report the mean with the half-width of the 95\% confidence interval.
R-Precision, FID, MultiModal Distance, Diversity, and MultiModality are computed using the standard evaluation feature extractor.
MultiModality uses 30 samples per caption and 10 sampled pairs. Baseline entries are the values published with each method, and a dash marks the methods whose papers do not report it.
R-Precision is reported at Top-1, Top-2, and Top-3.
FID and MultiModal Distance are computed in the evaluator feature space, and Diversity is computed by sampling pairs of generated motions following the standard protocol.
Real-motion rows are included in Table~\ref{tab:main_results} as references for interpreting FID, MultiModal Distance, and Diversity.
Real motion is not included when ranking generated methods.

\noindent\textbf{MotionMillion.}
MotionMillion results follow the MotionMillion benchmark protocol.
We use the MotionMillion evaluation split, retrieval protocol, and evaluation feature extractor for all reported MotionMillion numbers.

The ScaMo and MotionMillion-7B numbers are reported under the same MotionMillion evaluation protocol.

\subsection{Data-Derived Joint Groups}
\label{app:joint_groups}

Table~\ref{tab:joint_groups} lists the exact joint composition of the six data-derived groups used by the frozen tokenizer (Sec.~\ref{sec:method_tokenizer}), on the 22-joint HumanML3D skeleton.
The grouping is computed from motion statistics alone: per-joint activation statistics and joint--joint correlations at temporal lags up to 4 frames, estimated on $2\times10^5$ frames sampled from 23{,}368 training sequences, define a similarity matrix over joints; average-linkage hierarchical clustering proposes candidate groups; and a kinematic-chain correction reassigns joints whose chain would otherwise be split across groups.
Beyond the disjoint assignment, twelve boundary events copy the feature channels of a bridge joint into an adjacent group (e.g., pelvis and spine1 into both leg groups), so groups overlap at their kinematic boundaries and the six group encoders jointly cover all 263 HumanML3D feature channels.
Each group encoder receives the standard HumanML3D per-joint channels of its joints; the two leg groups additionally carry their foot-contact channels.
For KIT-ML the same procedure is applied to KIT's own motion statistics on its 21-joint, 251-channel representation, again with average-linkage clustering and $P=6$; the resulting grouping differs from the HumanML3D one, so the partition is re-derived per corpus rather than transferred.

\begin{table}[h]
\centering
\caption{Joint composition of the six data-derived groups.
Base joints come from the clustering assignment; boundary joints are shared with adjacent groups through channel duplication.
Group names describe dominant joints and are not hand-designed regions: the clustering separates the two legs, merges both arms with the upper spine, and isolates the head.}
\label{tab:joint_groups}
\small
\begin{tabular}{lll}
\toprule
Group & Base joints & Shared boundary joints \\
\midrule
root       & pelvis, spine1 & --- \\
upper arms & spine2, collars (l/r), shoulders (l/r), elbows (l/r), wrists (l/r) & spine1, neck \\
right leg  & right hip, knee, ankle, foot & pelvis, spine1, spine2 \\
upper neck & spine3, neck & spine2, collars (l/r) \\
left leg   & left hip, knee, ankle, foot & pelvis, spine1, spine2 \\
head       & head & neck \\
\bottomrule
\end{tabular}
\end{table}

\subsection{Geometry Diagnostic Protocol}
\label{app:geometry_protocol}

The metric-alignment diagnostic in Table~\ref{tab:geometry_diagnostics}(a) is computed on held-out HumanML3D motions using the frozen PartVQ tokenizer used by MoGeFlow.
Each group-specific codebook contains 128 entries.
We compute correlations over supported codes satisfying $|\Omega_{p,k}|\ge n_{\min}$.
For each supported code, we compute a local motion prototype using the subset of normalized motion features associated with the corresponding data-derived PartVQ group at the tokenizer time step.
A second, decoder-side prototype substitutes each code of a group into real contexts and decodes with the frozen decoder, averaging the decoded window at the substituted step restricted to that group's channels over 16 contexts per group.
The six groups are inherited from the frozen tokenizer and are descriptively named root, upper arms, right leg, upper neck, left leg, and head according to their dominant joints.
Embedding distance is Euclidean distance between frozen codebook entries.
Motion-prototype distance is Euclidean distance in the normalized local motion feature space.
The shuffled control randomly permutes the association between code embeddings and motion prototypes while preserving code usage statistics.

For the decoder-causal replacement diagnostic in Table~\ref{tab:motionmillion_and_replacement}(b), we replace a code in a valid code sequence with a near, mid-range, or far codebook neighbor.
We then decode the modified sequence with the frozen decoder and measure the change relative to the original decoded motion.
We report whole-body, target-group, and local target distances, all as mean absolute per-channel deviation from the unmodified decode.
The norm-matched control replaces the same code embedding by $e_{p,k}+\lVert e_{p,k'}-e_{p,k}\rVert\,u$ with $u$ uniform on the unit sphere, so the displacement has the same length but leaves the lattice; the panel reports 128 motions with one substitution per group.

\paragraph{Untrained-tokenizer control.}
We re-initialise the encoder and codebooks of the same architecture at random, re-encode the same motions with that tokenizer, build prototypes from its own assignments, and rerun the alignment diagnostic: the mean Spearman $\rho$ is $0.24$ against $0.822$ for the trained tokenizer, and only $37$ codes per group stay active against $113$. The correlation is insensitive to support size: subsampling the trained tokenizer to the same $37$ codes per group over 200 random draws leaves $\rho$ at $0.820$, against $0.822$ at full support.

\paragraph{Terminal off-lattice offset.}
On 512 prompts sampled with 96 Euler steps at guidance $6.0$, we measure for every valid frame and group the distance from the terminal state to its nearest code, normalized by half the median nearest-neighbor spacing of that group's codebook.
The pooled median is $0.68$, the mean $0.78$, and $26\%$ of frame--group states exceed $1$; per-group medians range from $0.61$ to $0.80$.

\paragraph{Standardized-metric check.}
The flow is trained on codebook-standardized latents (per-dimension mean/std of each group's codebook), so Table~\ref{tab:geometry_diagnostics}(a) reports the alignment correlations both in the raw code metric and under $D_p(k,k')=\|(e_{p,k}-e_{p,k'})\oslash\sigma_p\|_2$, computed on the same motions with 1000 bootstrap resamples.

\paragraph{Off-lattice offset directions.}
For the same terminal states we ask whether the off-lattice part of a state points along the local codebook directions.
For every valid frame and group we take the nearest code $e_{p,k^{*}}$, form the offset $\delta=y-e_{p,k^{*}}$, span the differences $e_{p,k_j}-e_{p,k^{*}}$ to the eight nearest codes, and report the fraction of $\lVert\delta\rVert^{2}$ lying in that subspace.
An isotropic direction would place $8/128=0.063$ of its energy there, and a norm-matched random control measures $0.063$.
The terminal offsets place $0.220$, a factor of $3.5$, with per-group factors from $2.7$ to $4.7$.
Terminal offsets are therefore concentrated along the directions that separate neighboring prototypes.

\paragraph{Generic residual-VQ control.}
Applying the identical protocol to the residual-VQ tokenizer used by the generic-RVQ ablation arms (six residual layers, 512 codes of dimension 512), with the layer axis in place of the part axis, gives learned $\rho=0.80$ [bootstrap 95\% CI $0.79,0.80$] for the base layer and $0.40$, $0.45$, $0.46$, $0.50$, $0.51$ for the five subsequent residual layers, with shuffled controls within $\pm0.07$.
Layer-wise replacement produces the same graded near/mid/far response only for the base layer.

\subsection{Ablation Protocol}
\label{app:ablation_protocol}

The ablations in Table~\ref{tab:key_ablations} are conducted on HumanML3D.
All variants use the same flow-only generation recipe unless otherwise specified, and are evaluated at guidance scale $6.0$; the flow arms integrate 96 Euler steps.
Every component taken from published code keeps that work's default settings, so the masked categorical arm decodes in MoMask's default 18 iterations.
The generic RVQ comparison changes the tokenizer/codebook interface while keeping the flow recipe fixed; its stage-grid variant treats the six residual levels as a per-step grid, while its summed-latent variant models the summed residual latent directly.
That tokenizer has six residual levels of 512 codes at dimension 512.
The model-capacity ablation changes the generator size while keeping the part-specific codebook interface fixed.
The masked categorical arm is a masked generative prior of the MoMask family trained over the same part-structured code lattice, predicting code indices under a cross-entropy objective instead of transporting states through code space.

The decode-mode rows of Table~\ref{tab:key_ablations} evaluate the same MoGeFlow-L states under both decoding paths.
The target-space comparison in Table~\ref{tab:geometry_diagnostics}(b) retrains the same flow with only the target changed: the encoder-latent arm regresses onto the same frozen tokenizer's pre-quantization encoder outputs, and the variational arm regresses onto the posterior mean of a PartVAE trained on the same data with the same partition and matched per-group latent width (128); that PartVAE is trained with a SmoothL1 reconstruction loss and a KL weight of $0.005$ annealed over the first $10^4$ iterations. Each target is standardized with its own per-dimension statistics.
Each comparison holds the flow recipe and the evaluation protocol fixed and changes one factor: the codebook interface, the generator size, the prior's objective, or the regression target.

\subsection{Implementation Details}
\label{app:implementation}

\begin{table}[h]
\centering
\caption{MoGeFlow variants. Hidden dimensions are part-hidden / global-hidden.}
\label{tab:model_configs}
\footnotesize
\begin{tabular}{lccccc}
\toprule
Variant & Params & Layers & Hidden dim & Heads & Dropout \\
\midrule
MoGeFlow-S & 173.38M & $6+12$ & $96/576$   & 12 & 0.05 \\
MoGeFlow-B & 307.55M & $6+12$ & $128/768$  & 12 & 0.05 \\
MoGeFlow-L & 690.47M & $6+12$ & $192/1152$ & 12 & 0.05 \\
\bottomrule
\end{tabular}
\end{table}

The frozen tokenizer is trained once with a SmoothL1 reconstruction loss, a local-joint-position term of weight $0.5$, and a commitment term of weight $0.02$; its codebooks are maintained by exponential moving averages of assigned encoder latents with dead-code reinitialization (decay $0.99$). It reconstructs held-out HumanML3D motions at FID $0.037$.
All variants (Table~\ref{tab:model_configs}) are trained with AdamW ($\beta=(0.9,0.95)$, base learning rate $1\times10^{-4}$, weight decay $0.01$), batch size $64$ for $600$ epochs, linear warmup of $2{,}000$ steps followed by half-cosine decay to $0.01\,\eta_0$, gradient clipping at global norm $1.0$, bfloat16 mixed precision, and condition dropout $0.1$ for classifier-free guidance; an EMA of the weights (decay $0.9999$) is used for evaluation, and the flow time is sampled uniformly.
Sampling uses Euler integration on a uniform grid with $96$ steps and guidance scale $6.0$ applied in velocity space, starting from $\mathcal{N}(0,I)$; terminal states are decoded directly by the frozen decoder.
All three datasets are trained for 600 epochs.

\end{document}